\documentclass[conference,a4paper]{IEEEtran}
\IEEEoverridecommandlockouts
\usepackage[T1]{fontenc}
\usepackage{cite}
\usepackage{amsmath,amssymb,amsfonts}
\usepackage{graphicx}
\usepackage{textcomp}
\usepackage{xcolor}
\usepackage{booktabs}
\usepackage{array}
\usepackage{tabularx}
\usepackage{stfloats} 
\usepackage[section,below]{placeins} 
\usepackage{url}
\usepackage[hidelinks]{hyperref}

\newif\ifshowtoc
\showtocfalse

\title{Quantifying Methodological Inflation in Clinical Machine Learning: A Cross-Modality Audit of Subject-Level Leakage and Nested Feature Selection for Parkinson's Disease}

\author{\IEEEauthorblockN{Namrata Datta}
\IEEEauthorblockA{\textit{Department of Mathematical Sciences} \\ \textit{Auckland University of Technology} \\ Auckland, New Zealand \\ dyt4883@autuni.ac.nz}
\and
\IEEEauthorblockN{Victor Miranda Soberanis}
\IEEEauthorblockA{\textit{Department of Mathematical Sciences} \\ \textit{Auckland University of Technology} \\ Auckland, New Zealand \\ victor.miranda@aut.ac.nz}}

\newcommand{\paperabstract}{%
\begin{abstract}
Reported accuracy in Parkinson's disease (PD) machine learning can reflect the evaluation protocol rather than real signal. We ask how much evaluation choices inflate performance by comparing naive and rigorous protocols on voice, telemonitoring and blood-transcriptomic datasets. Subject-aware splitting reduced voice AUROC from 0.892 to 0.780 and telemonitoring $R^2$ from 0.105 to $-0.037$; Random Forest recovered neither gap. Supervised probe selection before splitting inflated transcriptomic AUROC from 0.711 to 0.998. A nested 50-probe panel matched a 1,000-probe panel (0.811 versus 0.844), and recruitment-wave batch structure was detected. Future work should validate a stable panel on an independent cohort.
\end{abstract}

\begin{IEEEkeywords}
Parkinson's disease, data leakage, subject-level cross-validation, feature selection, transcriptomics, batch effects, corrected resampled $t$-test, reproducibility, explainable machine learning
\end{IEEEkeywords}
}

\begin{document}

\ifshowtoc
  \onecolumn
  \maketitle
  \paperabstract
  \clearpage
  \pagenumbering{roman}
  \setcounter{page}{1}
  {\footnotesize
  \tableofcontents
  \listoffigures
  \listoftables}
  \twocolumn
  \pagenumbering{arabic}
  \setcounter{page}{1}
\else
  \maketitle
  \paperabstract
\fi

\section{Introduction}\label{sec:intro}

Machine learning (ML) is increasingly explored as a low-cost, scalable route to detecting and monitoring Parkinson's disease (PD) from sustained-vowel acoustics \cite{little2007exploiting,little2009suitability}, at-home telemonitoring \cite{tsanas2010accurate}, blood transcriptomic profiles \cite{scherzer2007molecular,shamir2017analysis,calligaris2015blood}, multiplex blood and urine panels \cite{gao2025multiplex}, and multimodal imaging and clinical data from cohorts such as the Parkinson's Progression Markers Initiative (PPMI) \cite{dentamaro2024ppmi,makarious2022multimodality,zhang2022imaging}. Earlier diagnosis and objective monitoring would have clear clinical value, because motor symptoms typically appear only after substantial dopaminergic loss. The accuracy a model reports, however, depends heavily on how the data were split and processed. When the evaluation procedure allows information about the test data to reach the model, reported accuracy rises without any improvement in what the model would achieve on a new patient \cite{saeb2017need,kapoor2023leakage}. A recent systematic review of PD-ML studies found wide variation in cross-validation practice and patchy reporting of whether recordings from the same person were kept on one side of the split \cite{tabashum2024systematic}. \\

Two mechanisms of \emph{methodological inflation} are especially relevant to PD data. The first is \emph{subject-level leakage}: when each participant contributes several recordings, splitting by record rather than by person places near-duplicate observations of the same individual in both training and test sets, so a model can partly recognise people instead of disease \cite{saeb2017need,chaibubneto2019detecting}. The second is \emph{pre-split feature selection} in high-dimensional, small-sample ($p \gg n$) data such as transcriptomics, where choosing features with the outcome labels of the whole cohort lets the test labels shape the feature set \cite{ambroise2002selection,varma2006bias}. A third, less discussed problem concerns how protocols are compared: scores from repeated cross-validation are statistically dependent, and tests that treat them as independent overstate certainty \cite{nadeau2003inference}. \\

Although both leakage mechanisms are recognised in general terms, their size has rarely been measured side by side on structurally different PD datasets with the model held fixed. PD-specific evidence is limited to single-modality classification \cite{tougui2021impact,chaibubneto2019detecting}; no study has separated supervised from unsupervised pre-split selection in PD transcriptomics, and public PD transcriptomic cohorts have not been audited for recruitment structure using a test in which batch cannot act as a proxy for diagnosis. Closing these gaps matters because inflated estimates propagate into the literature, shape which biomarkers are pursued, and could ultimately lead to false reassurance or false alarm if translated into clinical tools. \\

The purpose of this study is to quantify how much of the performance reported for PD prediction models is attributable to the evaluation protocol rather than to generalisable signal, and to identify which safeguards matter for which data structure. The design is a controlled cross-modality audit of three public datasets: a 32-subject voice classification dataset, a 42-subject longitudinal telemonitoring regression dataset with roughly 140 recordings per subject, and a 59-subject blood transcriptomic dataset with 22,277 probes. Within each dataset the data, model family and hyper-parameter search are fixed and only the evaluation protocol varies. Elastic net is the primary model, and Random Forest is fitted on identical partitions to separate protocol effects from model capacity. The transcriptomic track adds supervised and unsupervised selection arms, a nested feature-efficiency analysis and a recruitment-wave audit, and all headline comparisons use inference procedures that respect the dependence between resampled estimates. \\

A \emph{naive protocol} is an evaluation procedure of the kind common in published PD-ML work, such as record-level splitting or feature selection on the full dataset. A \emph{rigorous protocol} is the corresponding procedure in which no information from held-out subjects reaches any fitted step. The \emph{inflation delta} is naive minus rigorous performance for the same model and task. A \emph{leakage-only ablation} is a comparison in which the two arms differ only in the leakage mechanism, with model, class weighting and number of folds identical. \emph{Fold-internal} selection means that feature ranking and scaling are fitted on the training fold alone.\\

The study addresses three research questions. The first asks whether subject-level leakage inflates performance consistently across structurally different repeated-measures PD datasets, and whether that inflation is attributable to the evaluation protocol rather than to limited model capacity. The second asks whether pre-split feature selection inflates performance in a $p \gg n$ transcriptomic cohort, whether this depends on the selection being supervised, and whether a compact panel selected entirely within training folds can match a large panel. The third asks whether the transcriptomic diagnostic signal is accompanied by non-biological recruitment structure, and whether the signal persists once that structure is balanced out.

The paper makes five contributions:

\begin{itemize}
\item to our knowledge, the first controlled three-modality leakage audit for PD that spans classification, regression and high-dimensional transcriptomics, with the model family held fixed and a leakage-only ablation;
\item a direct separation of supervised from unsupervised pre-split selection in PD transcriptomics, showing that the former drives a chance-level model to near-perfect apparent accuracy while the latter has no effect;
\item a nested, leakage-free feature-efficiency analysis in which a 50-probe panel matches the 1,000-probe reference, together with per-probe selection stability across folds;
\item the identification of recruitment-wave batch structure in the public GSE72267 cohort, using a within-patient test that cannot be explained by the outcome;
\item a practical inference framework for protocol comparisons, including the observation that row shuffling does not re-randomise scikit-learn's \texttt{GroupKFold}, and a paired subject-level bootstrap that makes record-level and subject-level protocols directly comparable.
\end{itemize}

Section \ref{sec:related} reviews related work. Section \ref{sec:rq} states the aims, hypotheses and scope. Section \ref{sec:methods} describes the datasets, protocols, models and statistical procedures. Section \ref{sec:results} reports the results, Section \ref{sec:discussion} discusses them, Section \ref{sec:evaluation} evaluates the research method, and Section \ref{sec:conclusion} concludes.

\section{Literature Review and Related Work}\label{sec:related}

\subsection{Acoustic and Telemonitoring Biomarkers}

Little et al. \cite{little2007exploiting} introduced non-linear recurrence and fractal-scaling measures for voice disorder detection and later applied them to the 32-subject UCI Parkinson's voice dataset, reporting around 91\% classification accuracy from features such as pitch period entropy (PPE), recurrence period density entropy (RPDE) and detrended fluctuation analysis (DFA) \cite{little2009suitability,little_uci_2007}. Sakar et al. \cite{sakar2013collection} extended this work to a larger cohort with multiple recording types, and Tsanas et al. \cite{tsanas2010accurate} showed that at-home voice tests could track Unified Parkinson's Disease Rating Scale (UPDRS) severity over six months. Recent vocal-biomarker work applies explainable ML with SHAP attribution to early detection in ageing populations \cite{egbo2025vocal}. A recurring feature of the follow-on literature using these datasets is random record-level cross-validation, even though each participant contributes six or seven (voice) or about 140 (telemonitoring) highly correlated recordings. Tougui et al. \cite{tougui2021impact} showed on PD voice data that record-wise cross-validation yields markedly higher accuracy than subject-wise cross-validation, and Chaibub Neto et al. \cite{chaibubneto2019detecting} showed on smartphone PD data that record-wise splits allow models to exploit subject identity. These studies establish the phenomenon for classification; they do not extend it to severity regression, to high-dimensional modalities, or to dependence-aware inference.

\subsection{Blood and Multiplex Biomarkers}

Scherzer et al. \cite{scherzer2007molecular} reported a blood expression signature for early PD, and Shamir et al. \cite{shamir2017analysis} analysed blood expression across several cohorts, reporting more modest discrimination when signatures were tested beyond their discovery data. GSE72267 was described by Calligaris et al. \cite{calligaris2015blood} and deposited in GEO \cite{roncaglia2015geo}; the authors highlighted chromatin-remodelling and immune pathways in drug-na\"{\i}ve patients. These cohorts share the defining difficulty of transcriptomic ML: tens of subjects and tens of thousands of features. Ambroise and McLachlan \cite{ambroise2002selection} showed that selecting genes with the full labelled dataset before cross-validation can make random data appear highly predictive, and Varma and Simon \cite{varma2006bias} showed similar optimism when tuning is not nested. Technical batch structure is a separate threat in small multi-batch cohorts and can masquerade as biology \cite{leek2010tackling,johnson2007adjusting}. Recent clinical blood-panel models \cite{han2025blood} and multiplex blood and urine panels \cite{gao2025multiplex} pursue compact, deployable markers, which makes leakage-free panel validation and batch auditing directly relevant to translation.

\subsection{Multimodal and Deep-Learning Approaches}

Work on PPMI combines imaging, clinical, biospecimen and digital measures. Dentamaro et al. \cite{dentamaro2024ppmi} apply multimodal deep learning with explainable-AI attribution, Makarious et al. \cite{makarious2022multimodality} combine genetic, transcriptomic and clinical data, and Zhang \cite{zhang2022imaging} reviews imaging and clinical ML for diagnosis. These studies demonstrate the value of fusion and explainability. As Tabashum et al. \cite{tabashum2024systematic} note, however, PD-ML studies rarely report protocol ablations, so it is not possible to tell from published metrics how much accuracy would survive subject-aware or fold-internal evaluation.

\subsection{Leakage, Model Selection and Inference in Clinical ML}

Saeb et al. \cite{saeb2017need} argue that the cross-validation scheme must mimic the intended use and show that record-wise splits overestimate subject-level accuracy in digital-health data. Kapoor and Narayanan \cite{kapoor2023leakage} catalogue leakage across 17 scientific fields and identify pre-processing on the full dataset and train-test dependence as the most common forms. Cawley and Talbot \cite{cawley2010overfitting} show that model selection on the evaluation data biases performance estimates, and Vabalas et al. \cite{vabalas2019machine} and Varoquaux \cite{varoquaux2018cross} show that in small samples cross-validated estimates carry large error bars. On inference, Nadeau and Bengio \cite{nadeau2003inference} show that naive $t$-tests on resampled fold scores are anti-conservative because training sets overlap, and propose a corrected variance estimator. Claims that two methods perform equivalently require a pre-specified margin and one-sided tests \cite{schuirmann1987comparison}.

\subsection{Synthesis and Research Gaps}

Three gaps emerge from this literature. First, PD-specific leakage evidence is confined to single-modality classification, and no study has shown in one framework that the effect extends to severity regression while ruling out model capacity as the explanation. Second, the PD transcriptomic literature has not distinguished unsupervised pre-filtering, which uses no labels, from supervised selection, which does; the two are often conflated in methods sections even though only the latter can leak test labels. Third, public PD transcriptomic cohorts have not been audited for recruitment or batch structure with a design in which the batch variable is not a proxy for the outcome. Across all three, protocol comparisons commonly rely on tests that ignore fold dependence. Table \ref{tab:related} positions representative studies against these gaps.

\section{Study Objectives and Scope}\label{sec:rq}

The aim is to quantify protocol-driven performance inflation in PD machine learning across structurally different data and to establish which evaluation safeguards each data structure requires. The objectives are to measure the inflation produced by record-level splitting in a classification and a regression dataset with repeated measurements; to test whether a higher-capacity model recovers the performance lost under subject-aware evaluation; to measure the inflation produced by supervised and unsupervised pre-split selection in a transcriptomic cohort; to assess whether a compact, fold-internally selected probe panel retains the performance of a large panel; to test the transcriptomic signal for non-biological recruitment structure; and to compare protocols with inference that respects the dependence structure of resampled estimates.

\begin{table*}[!tb]
\centering
\caption{Representative prior Parkinson's disease ML studies and their relation to this work}
\label{tab:related}
\small
\renewcommand{\arraystretch}{1.15}
\begin{tabularx}{\linewidth}{p{2.6cm}p{2.7cm}p{2.4cm}X}
\toprule
\textbf{Study} & \textbf{Modality / data} & \textbf{Subject-aware split} & \textbf{Relation to this work} \\
\midrule
Little et al. \cite{little2009suitability} & Voice (UCI, 32 subjects) & Not reported & Same dataset; subject-aware evaluation reduces AUROC substantially while leaving genuine above-chance signal. \\
Sakar et al. \cite{sakar2013collection} & Voice (extended) & Record-level & Same repeated-recording structure; leakage not addressed. \\
Tougui et al. \cite{tougui2021impact} & Voice (PD) & Compares both & Closest classification precedent; extended here to regression, transcriptomics and dependence-aware tests. \\
Chaibub Neto et al. \cite{chaibubneto2019detecting} & Smartphone (mPower) & Compares both & Identity confounding in PD digital data; motivates the subject-level design. \\
Tsanas et al. \cite{tsanas2010accurate} & Telemonitoring, UPDRS & Record-level & Same dataset; under subject-aware splits acoustic features do not exceed a mean-prediction baseline. \\
Egbo et al. \cite{egbo2025vocal} & Vocal biomarkers + SHAP & Not explicit & The audit framework quantifies the leakage such pipelines must rule out. \\
Scherzer et al. \cite{scherzer2007molecular} & Blood expression & Not reported & Supervised full-data selection shown here to inflate AUROC in a comparable cohort. \\
Shamir et al. \cite{shamir2017analysis} & Blood expression, multi-cohort & Cross-cohort & No batch audit reported; recruitment structure identified here in GSE72267. \\
Han et al. \cite{han2025blood}; Gao et al. \cite{gao2025multiplex} & Clinical / multiplex blood & Not explicit & Compact-panel aims match ours; the nested design is a validation template. \\
Dentamaro et al. \cite{dentamaro2024ppmi}; Makarious et al. \cite{makarious2022multimodality} & Multimodal (PPMI) & Not an ablation & Natural targets for the audit framework. \\
Saeb et al. \cite{saeb2017need}; Kapoor and Narayanan \cite{kapoor2023leakage} & Methodological & Yes & General leakage frameworks, applied here to three PD modalities. \\
\bottomrule
\end{tabularx}
\end{table*}

The study focuses on the research questions below:
\begin{itemize}
    \item The first research question follows directly from the first gap: does subject-level leakage inflate performance consistently across repeated-measures PD datasets, and is the inflation attributable to the protocol rather than to model capacity? We hypothesised that record-level evaluation would yield higher performance than subject-level evaluation in both the voice and the telemonitoring datasets, and that a Random Forest evaluated on the same subject-level partitions would not close the gap.

\item The second research question addresses the transcriptomic gap and has two parts. The first part asks whether pre-split feature selection inflates performance and whether this depends on the selection being supervised; we hypothesised that supervised selection on the full cohort would inflate AUROC whereas unsupervised variance filtering would not. The second part asks whether a compact panel chosen entirely within training folds can match a large panel; we hypothesised that importance-based nested selection would allow a 50-probe panel to perform comparably to the 1,000-probe reference, and that selection strategy would matter more than panel size.

\item The third research question concerns cohort structure: is the transcriptomic signal accompanied by non-biological recruitment structure, and does diagnostic signal persist when that structure is balanced? This study hypothesised that expression would predict recruitment wave even among patients who share the same diagnosis, and that diagnostic AUROC would remain clearly above chance in a wave-balanced matched-pair subset.
\end{itemize}

The study is an internal-validation, protocol-comparison analysis of three public datasets. Its object of study is the evaluation procedure rather than a new predictive architecture, so deep-learning models, external-cohort validation, batch-correction methods such as ComBat \cite{johnson2007adjusting} and pathway-enrichment analysis lie outside its scope; selected probes are annotated to gene symbols. The datasets share no participants, so each is analysed as an independent track under a common framework.

\section{Methodology}\label{sec:methods}

\subsection{Design Rationale and Alternatives Considered}

The study uses a within-dataset controlled comparison: for each dataset the data, model family and hyper-parameter search are fixed and only the evaluation protocol varies. This isolates the protocol effect far better than comparing published numbers across papers, which differ in many uncontrolled ways. Several alternative methods from the statistical, ML and bioinformatics literature were evaluated before the design was fixed (Table \ref{tab:alternatives}).

\begin{table*}[!tb]
\centering
\caption{Methodological alternatives considered}
\label{tab:alternatives}
\small
\renewcommand{\arraystretch}{1.15}
\begin{tabularx}{\linewidth}{p{3.8cm}p{2.0cm}X}
\toprule
\textbf{Method} & \textbf{Decision} & \textbf{Justification} \\
\midrule
Leave-one-subject-out CV & Not primary & With 8 voice controls, per-fold AUROC is undefined for most folds; pooled out-of-fold predictions provide the same subject-level guarantee with stable estimates. \\
Repeated subject-level K-fold & Adopted & Yields many distinct subject partitions when implemented with StratifiedGroupKFold and explicit shuffling. \\
DeLong test on a single split & Replaced & Requires matched predictions; a paired subject-level bootstrap of pooled out-of-fold predictions achieves pairing across protocols. \\
Mann--Whitney / Wilcoxon on fold scores & Rejected & Treats overlapping-fold scores as independent and is anti-conservative \cite{nadeau2003inference}. \\
Corrected resampled $t$-test & Adopted & Standard variance correction for repeated cross-validation \cite{nadeau2003inference}. \\
Deep learning models & Rejected & 32--59 subjects per dataset; the question concerns protocols, not architectures. \\
ComBat batch correction & Deferred & Batch labels are inferred, not recorded, and correction would itself require nesting; structure is audited instead. \\
\bottomrule
\end{tabularx}
\end{table*}

\subsection{Datasets}

Three public datasets were analysed as separate tracks (Table \ref{tab:datasets}). The voice track uses the UCI Parkinson's dataset \cite{little_uci_2007,little2009suitability}; subject identity was recovered from the \texttt{name} field (for example \texttt{phon\_R01\_S01\_1}). The telemonitoring (UPDRS) track uses the UCI telemonitoring dataset \cite{tsanas2010accurate} with subject identifier \texttt{subject\#}; the target was total UPDRS, and \texttt{motor\_UPDRS} and \texttt{test\_time} were excluded as predictors. The blood track uses GEO series GSE72267 \cite{calligaris2015blood,roncaglia2015geo}, profiled on the Affymetrix HG-U133A 2.0 array (platform GPL571).

\begin{table*}[!tb]
\centering
\caption{Dataset characteristics}
\label{tab:datasets}
\small
\renewcommand{\arraystretch}{1.15}
\begin{tabularx}{\linewidth}{p{3.4cm}XXX}
\toprule
 & \textbf{Voice} & \textbf{UPDRS} & \textbf{Blood} \\
\midrule
Source & UCI Parkinsons & UCI Telemonitoring & GEO GSE72267 \\
Subjects & 32 (24 PD, 8 control) & 42 (all PD) & 59 (40 PD, 19 control) \\
Observations & 195 recordings & 5,875 recordings & 59 arrays \\
Observations per subject & 6--7 & about 140 & 1 \\
Features & 22 acoustic & 16 acoustic, age, sex & 22,277 probes \\
Task & Binary classification & Regression (total UPDRS) & Binary classification \\
Principal leakage risk & Subject identity & Subject identity & Feature selection; batch \\
\bottomrule
\end{tabularx}
\end{table*}

\subsection{Evaluation Protocols}

For each track a naive protocol, representative of common practice, was compared with a rigorous protocol (Table \ref{tab:protocols}). For voice, a leakage-only ablation was also run in which both arms use class-weighted elastic net and five folds and differ only in whether folds respect subject identity (record-level StratifiedKFold versus StratifiedGroupKFold, each with ten shuffled repeats). For UPDRS the naive and rigorous protocols differ only in grouping. For blood, two leakage arms were evaluated on identical folds with identical class weighting: an unsupervised arm, ranking probes by variance, and a supervised arm, ranking probes by the absolute Welch $t$-statistic between PD and controls. Each ranking was computed either on the full cohort before splitting or inside each training fold.

\begin{table*}[!tb]
\centering
\caption{Naive and rigorous protocol definitions}
\label{tab:protocols}
\small
\renewcommand{\arraystretch}{1.15}
\begin{tabularx}{\linewidth}{p{2.6cm}XX}
\toprule
\textbf{Track} & \textbf{Naive protocol} & \textbf{Rigorous protocol} \\
\midrule
Voice (headline) & Random 70/30 record-level split, 20 repeats, no class weighting & GroupKFold (5 folds) by subject, class-weighted \\
Voice (ablation) & Record-level StratifiedKFold, 5 folds $\times$ 10 repeats, class-weighted & StratifiedGroupKFold, 5 folds $\times$ 10 shuffled repeats, class-weighted \\
UPDRS & Record-level KFold, 5 folds $\times$ 3 repeats & GroupKFold by \texttt{subject\#}; 5 random subject partitions for inference \\
Blood (unsupervised) & Variance filter on full cohort, then repeated stratified 5-fold & Variance filter and scaler refitted inside each training fold \\
Blood (supervised) & t-statistic ranking on full cohort, then repeated stratified 5-fold & t-statistic ranking inside each training fold \\
\bottomrule
\end{tabularx}
\end{table*}

\subsection{Predictive Models}

The primary model in every track is elastic-net regularised: $\ell_1/\ell_2$-penalised logistic regression for classification and \texttt{ElasticNetCV} for regression, minimising

\begin{equation}
\mathcal{L}(\beta) = \mathcal{L}_0(\beta) + \lambda \left[ \rho \lVert \beta \rVert_1 + \tfrac{1-\rho}{2}\lVert \beta \rVert_2^2 \right]
\label{eq:en}
\end{equation}

where $\mathcal{L}_0$ is the log-loss or squared error. The penalty strength $\lambda$ and mixing ratio $\rho$ were tuned by an inner three-fold cross-validation within each training fold ($\rho \in \{0.3, 0.5, 0.7\}$ for classification and $\{0.1, \dots, 0.9\}$ for regression), and features were standardised with training-fold statistics only. The $\ell_1$ term performs automatic feature selection, which is important in the $p \gg n$ blood setting. A Random Forest (500 trees, minimum leaf size 2, balanced class weights for classification) was fitted on the identical rigorous partitions, so that differences between the two models reflect model form rather than splitting.

\subsection{Repeated Subject-Level Partitioning}

Repeated cross-validation is commonly implemented by shuffling rows before re-applying a splitter. For scikit-learn's \texttt{GroupKFold} this does not work: groups are assigned to folds deterministically by size, so every shuffled repeat reproduces the same subject partition. We verified this empirically (ten row permutations of the voice data produced identical subject-to-fold assignments). Repeated subject-level estimates were therefore generated with \texttt{StratifiedGroupKFold(shuffle=True)} for voice, which produced ten distinct partitions, and with seeded random subject-to-fold assignment for UPDRS.

\subsection{Feature Stability and Interpretation}

Feature stability was estimated as the proportion of bootstrap refits (100 for voice, 50 for UPDRS) in which an elastic-net coefficient was non-zero. SHAP values \cite{lundberg2017unified} were computed with a linear explainer for the rigorous elastic-net model, following interpretability practice in PD vocal-biomarker and multimodal studies \cite{egbo2025vocal,dentamaro2024ppmi}. Both analyses are used descriptively, alongside rather than in place of the predictive evaluation.

\subsection{Blood Feature-Reduction Analyses}\label{sec:bloodmethods}

Two analyses were run with repeated stratified 5-fold cross-validation (3 repeats, 15 folds), with all selection confined to training folds. The panel-size sweep evaluated variance-selected panels of $k \in \{10, 25, 50, 100, 250, 500, 1000, 2000\}$ probes with elastic net and Random Forest. Nested importance selection then fitted, within each training fold, a Random Forest to the 1,000 highest-variance training probes, kept the top $k \in \{10, 25, 50\}$ by impurity importance, and fitted a fresh classifier on those probes alone before scoring the held-out fold. Because the selected probes may differ between folds, the object validated is the selection procedure; panel composition is described by each probe's selection frequency across the 15 folds, and probes were annotated with GPL571 gene symbols.

\subsection{Recruitment-Wave Audit}

GSE72267 has no recorded batch field, but the raw CEL file names carry an enrolment code. Grouping samples by that code reveals 36 samples in codes containing both a PD patient and a control (the \emph{matched-pair wave}; 18 PD, 18 controls) and 23 samples in codes containing only one class (the \emph{unpaired wave}; 22 PD, 1 control). Because wave and diagnosis are strongly associated in the full cohort, predicting wave in the full cohort cannot distinguish biology from batch. The audit therefore predicts wave \emph{among PD patients only} (18 versus 22), where wave cannot act as a proxy for diagnosis, using a Random Forest on 1,000 fold-internal variance probes with a 100-permutation label-shuffling null. Diagnostic performance was also estimated in the wave-balanced matched-pair subset (repeated stratified 4-fold, 5 repeats).

\subsection{Statistical Inference}\label{sec:stats}

Scores from repeated cross-validation are not independent because training sets overlap, so the sample variance of fold scores understates the variance of their mean. Headline comparisons therefore used three procedures matched to the dependence structure of each comparison:

\begin{itemize}
\item \textbf{Different partitions} (naive versus rigorous, voice and UPDRS): a Welch-type statistic in which each arm's variance is inflated by the Nadeau--Bengio correction $\hat\sigma^2$(1/J + $n_{\mathrm{test}}/n_{\mathrm{train}}$), where J is the number of fold scores \cite{nadeau2003inference}.
\item \textbf{Same records, different predictions} (voice and UPDRS): because both protocols produce an out-of-fold prediction for every record, a paired subject-level cluster bootstrap (2,000 resamples for voice, 1,000 for UPDRS) of the difference in pooled AUROC or $R^2$, resampling whole subjects.
\item \textbf{Same folds} (blood): the corrected resampled paired $t$-test \cite{nadeau2003inference}, with a one-sided non-inferiority test of the 50-probe against the 1,000-probe panel at margins of 0.05 and 0.10 AUROC \cite{schuirmann1987comparison}.
\end{itemize}

Confidence intervals are 95\% intervals from the same procedures and describe uncertainty within each dataset. The three primary comparisons (pooled voice AUROC, pooled UPDRS $R^2$ and supervised selection) address separate pre-specified hypotheses; they remain significant after Holm correction across the five naive-versus-rigorous tests, and the two fold-level tests are reported as supporting analyses.

\subsection{Ethical Considerations}

All three datasets are public and de-identified, were collected under the ethics approvals of the original studies \cite{little2009suitability,tsanas2010accurate,calligaris2015blood}, and are distributed for research reuse; no new data were collected and no re-identification was attempted. The principal ethical risk in this area is over-claiming: a PD classifier with inflated accuracy could produce false reassurance or false alarm if translated into practice. The study therefore reports rigorous estimates as primary, quantifies uncertainty for every headline comparison, and presents the compact transcriptomic panel as a validated selection procedure rather than a diagnostic test. Class imbalance (8 voice controls; a single control in the unpaired blood wave) is reported explicitly because it bounds how far conclusions extend to healthy populations.

\subsection{Implementation and Reproducibility}

\begin{figure}[!htbp]
\centering
\includegraphics[width=0.92\linewidth]{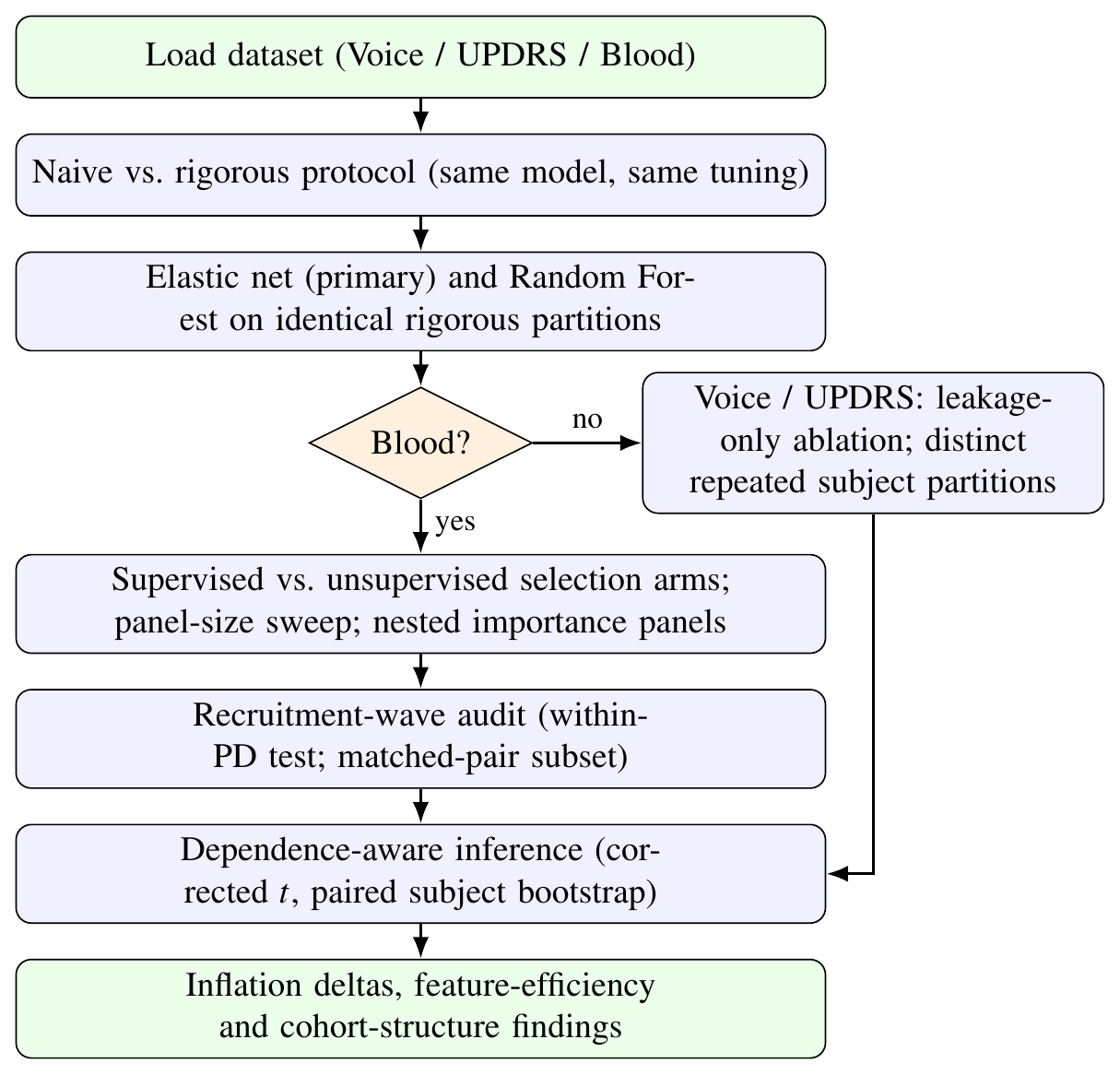}
\caption{Analysis workflow. Voice and UPDRS follow the subject-level leakage branch; the blood track adds selection, feature-efficiency and recruitment-wave analyses. All tracks share the dependence-aware inference step.}
\label{fig:flow}
\end{figure}

Analyses used Python with scikit-learn 1.6, SciPy, pandas and SHAP, with a fixed random seed of 42. Each pipeline writes every reported quantity to CSV files, and the numbers, tables and figures in this paper are generated programmatically from those files. Single-partition GroupKFold estimates depend on how ties between equal-sized subjects are ordered, which can differ across hardware (voice AUROC 0.76--0.81 across tie orders); the repeated StratifiedGroupKFold estimates, which are hardware-independent, are therefore used for all inference.
Fig. \ref{fig:flow} summarises the analysis workflow.

\section{Results}\label{sec:results}

\subsection{Voice: Subject-Level Leakage Inflates Classification Performance}

\begin{figure}[!htbp]
\centering
\includegraphics[width=1.0\linewidth]{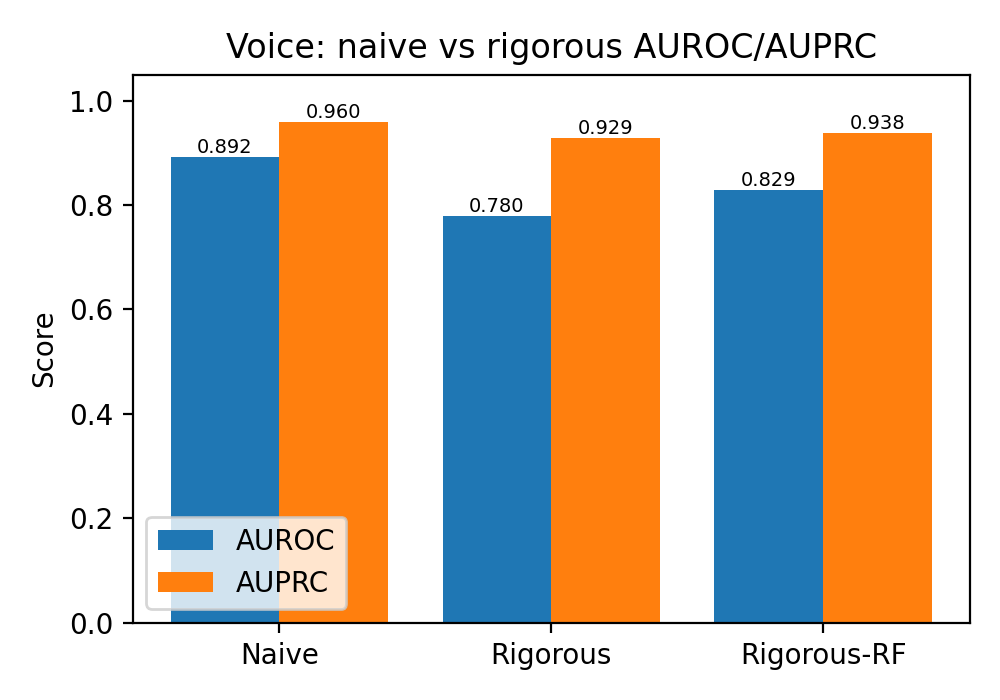}
\caption{Voice track: AUROC and AUPRC under the naive protocol, the rigorous protocol and Random Forest on the rigorous partitions.}
\label{fig:voice}
\end{figure}

Table \ref{tab:voice} and Fig. \ref{fig:voice} report the headline protocols. Record-level evaluation gave AUROC 0.892, while subject-aware evaluation gave 0.780, a drop of 0.112. AUPRC fell from 0.960 to 0.929; AUPRC remains high under both protocols because 75\% of subjects have PD. A Random Forest on the same subject-aware folds reached 0.829, recovering part but not all of the gap.

\begin{table}[!htbp]
\centering
\caption{Voice track: naive versus rigorous performance (mean $\pm$ SD)}
\label{tab:voice}
\small
\renewcommand{\arraystretch}{1.15}
\begin{tabularx}{\linewidth}{Xcc}
\toprule
\textbf{Protocol} & \textbf{AUROC} & \textbf{AUPRC} \\
\midrule
Naive (record-level 70/30, 20 repeats) & 0.892 $\pm$ 0.046 & 0.960 $\pm$ 0.019 \\
Rigorous (GroupKFold by subject) & 0.780 $\pm$ 0.132 & 0.929 $\pm$ 0.041 \\
Rigorous, Random Forest & 0.829 $\pm$ 0.120 & 0.938 $\pm$ 0.055 \\
Ablation: record-level 5-fold $\times$ 10 (pooled) & 0.892 & -- \\
Ablation: subject-level 5-fold $\times$ 10 (pooled) & 0.693 & -- \\
\bottomrule
\end{tabularx}
\end{table}

The leakage-only ablation confirms that grouping alone drives the effect. Across ten distinct subject partitions, pooled out-of-fold AUROC was 0.892 for record-level folds and 0.693 for subject-level folds (SD across repeats 0.063). The paired subject-level bootstrap gave a difference of 0.200 (95\% CI [0.087, 0.330], one-sided p $<$ 0.001). Mean fold-level AUROC showed the same gap under the corrected resampled $t$-test (0.899 versus 0.723; difference 0.175, 95\% CI [$-$0.021, 0.371], corrected $p$ = 0.039). Pooled estimates are reported alongside fold means because per-fold AUROC is unstable when a subject-level test fold contains only one or two controls.

The rigorous elastic-net coefficients and SHAP values ranked MDVP:APQ, spread2, DFA, D2, MDVP:PPQ and PPE highest (Table \ref{tab:voicefeat}), while bootstrap stability ranked spread1, PPE and spread2 highest (selection frequency $\geq$ 0.98 under both protocols). The amplitude-perturbation and non-linear dysphonia measures are strongly correlated, so elastic net distributes weight among interchangeable features; the stable core of the signal lies in this correlated family of PPE, spread and perturbation measures rather than in any single feature.

\begin{table}[!htbp]
\centering
\caption{Voice track: top features of the rigorous elastic-net model}
\label{tab:voicefeat}
\small
\renewcommand{\arraystretch}{1.15}
\begin{tabular}{lcc}
\toprule
\textbf{Feature} & \textbf{$|$Coefficient$|$} & \textbf{Mean $|$SHAP$|$} \\
\midrule
MDVP:APQ & 1.699 & 1.169 \\
spread2 & 1.418 & 1.123 \\
MDVP:PPQ & 0.788 & 0.487 \\
DFA & 0.771 & 0.666 \\
D2 & 0.617 & 0.489 \\
PPE & 0.420 & 0.334 \\
\bottomrule
\end{tabular}
\end{table}

\subsection{UPDRS: Repeated-Measures Leakage Produces the Most Severe Inflation}

\begin{figure}[!htbp]
\centering
\includegraphics[width=1.0\linewidth]{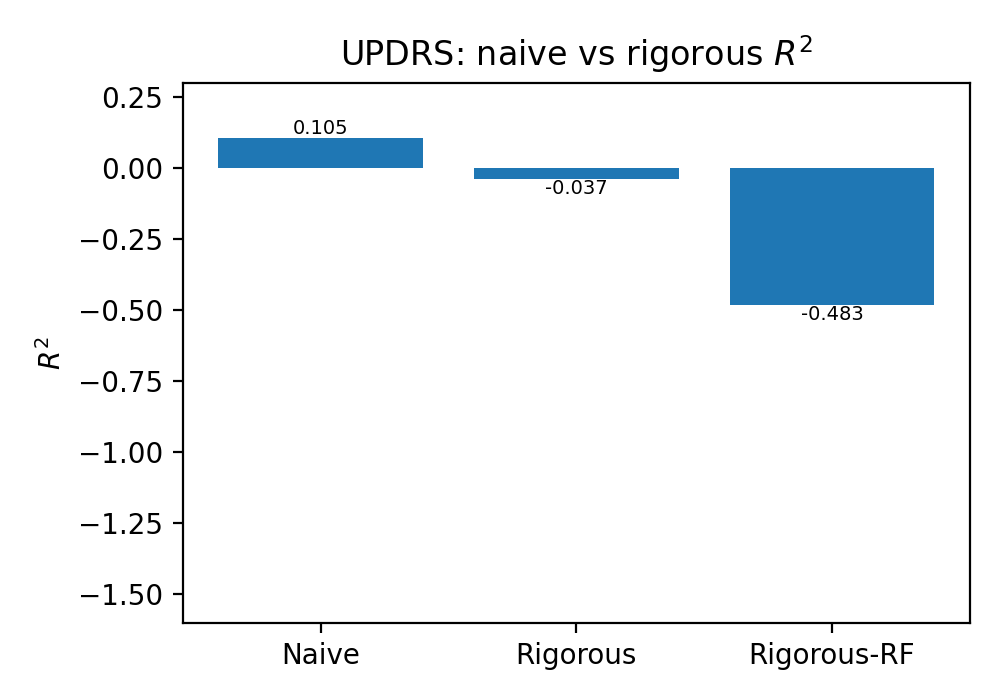}
\caption{UPDRS track: $R^2$ under the naive protocol, the rigorous protocol and Random Forest on the rigorous partitions; both rigorous estimates fall below the zero-skill line.}
\label{fig:updrs}
\end{figure}

Record-level evaluation gave $R^2$ = 0.105, whereas subject-level GroupKFold gave $R^2$ = $-$0.037, and a Random Forest on the same folds gave $-$0.483 (Table \ref{tab:updrs}, Fig. \ref{fig:updrs}). Across five random subject partitions, the paired subject-level bootstrap of pooled out-of-fold $R^2$ gave a difference of 0.171 (95\% CI [0.138, 0.225], p $<$ 0.001). Under subject-level evaluation, mean absolute error (MAE) was 9.12 points against 8.91 for a model predicting the training-fold mean (corrected paired $p$ = 0.716), so acoustic features provide no measurable improvement over a constant prediction for an unseen patient. The non-linear model performed worse rather than better, which rules out insufficient linear capacity as an explanation for the negative $R^2$.

\begin{table}[!htbp]
\centering
\caption{UPDRS track: naive versus rigorous performance (mean $\pm$ SD)}
\label{tab:updrs}
\small
\renewcommand{\arraystretch}{1.15}
\begin{tabularx}{\linewidth}{Xcc}
\toprule
\textbf{Protocol} & \textbf{$R^2$} & \textbf{MAE (points)} \\
\midrule
Naive (record-level KFold) & 0.105 $\pm$ 0.006 & 8.30 $\pm$ 0.16 \\
Rigorous (GroupKFold) & $-$0.037 $\pm$ 0.085 & 8.86 $\pm$ 0.67 \\
Rigorous, Random Forest & $-$0.483 $\pm$ 0.337 & 10.53 $\pm$ 1.06 \\
Training-mean baseline (subject-level) & 0 by definition & 8.91 \\
\bottomrule
\end{tabularx}
\end{table}

SHAP attributed most of the rigorous model's prediction to age (mean $|$SHAP$|$ = 0.805) and DFA (0.453), with RPDE and PPE contributing little and other features shrunk to zero. Because age varies between subjects but not within them, this pattern indicates that the record-level model's apparent skill came largely from learning who each patient was rather than from an acoustic marker of severity.

\subsection{Blood: Supervised, Not Unsupervised, Pre-Split Selection Drives Inflation}

\begin{figure*}[!tb]
\centering
\includegraphics[width=0.72\linewidth]{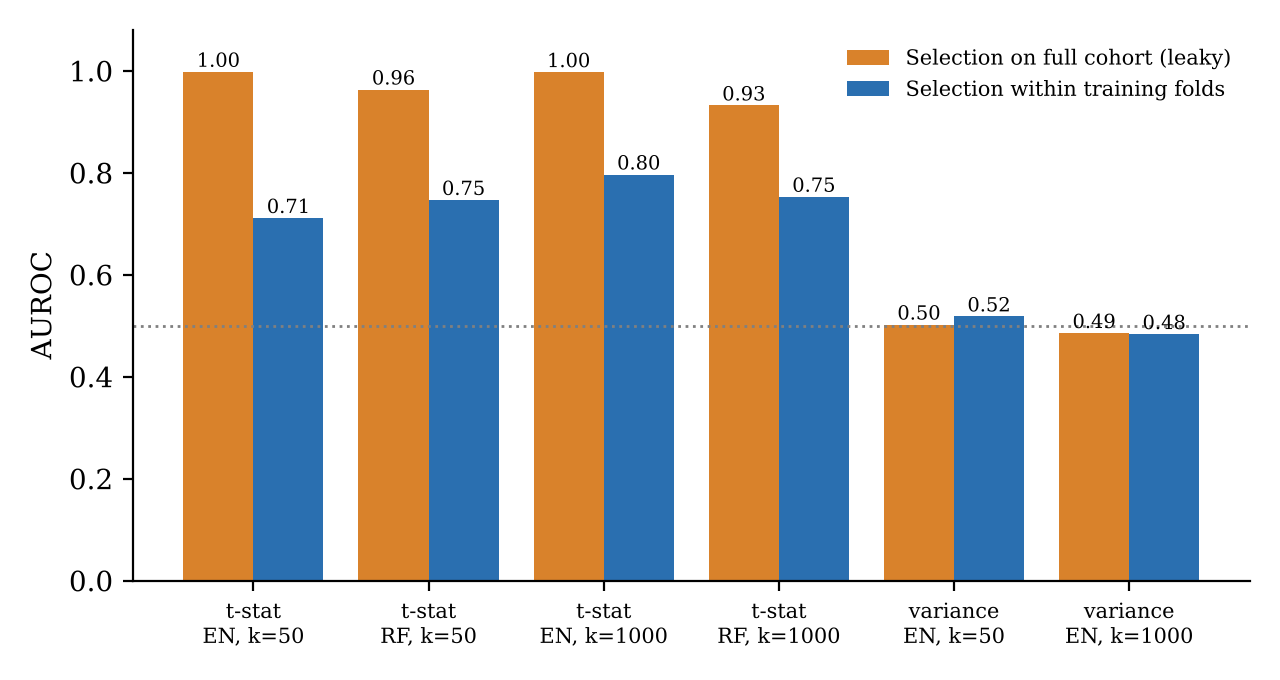}
\caption{Blood track: AUROC with feature selection on the full cohort (leaky) versus within training folds. Supervised (t-statistic) selection on the full cohort inflates performance; unsupervised variance filtering does not.}
\label{fig:leak}
\end{figure*}

With 1,000 variance-filtered probes, elastic net performed at chance under both protocols (AUROC 0.502 naive versus 0.476 rigorous); inspection of the full-data fit showed that the penalty shrank every coefficient to zero, reducing the classifier to an intercept. Table \ref{tab:leak} and Fig. \ref{fig:leak} isolate the selection step. Unsupervised variance filtering on the full cohort produced no inflation at either panel size. Supervised $t$-statistic ranking on the full cohort produced large inflation: with 50 probes, elastic-net AUROC was 0.998 under full-cohort selection against 0.711 under fold-internal selection (difference 0.287, 95\% CI [0.096, 0.477], $p$ = 0.003), and the effect was of similar size for Random Forest and at $k=$ 1,000. Supervised selection performed within training folds also turned a chance-level elastic net into a genuinely informative model (AUROC 0.71--0.80), showing that label-informed selection is valuable when it is kept inside the training data.

\begin{table*}[!tb]
\centering
\caption{Blood track: AUROC with selection on the full cohort versus within training folds (identical folds, model and class weighting; corrected paired $t$-test, one-sided)}
\label{tab:leak}
\footnotesize
\renewcommand{\arraystretch}{1.15}
\begin{tabularx}{\linewidth}{llcccXc}
\toprule
\textbf{Selection} & \textbf{Model} & \textbf{k} & \textbf{Full cohort} & \textbf{In fold} & \textbf{Difference [95\% CI]} & \textbf{p} \\
\midrule
Supervised ($t$) & EN & 50 & 0.998 & 0.711 & 0.287 [0.096, 0.477] & 0.003 \\
Supervised ($t$) & RF & 50 & 0.963 & 0.747 & 0.217 [0.031, 0.402] & 0.013 \\
Supervised ($t$) & EN & 1,000 & 0.997 & 0.797 & 0.201 [$-$0.001, 0.403] & 0.026 \\
Supervised ($t$) & RF & 1,000 & 0.932 & 0.753 & 0.179 [$-$0.009, 0.367] & 0.030 \\
Variance & EN & 50 & 0.502 & 0.518 & $-$0.016 [$-$0.081, 0.049] & 0.698 \\
Variance & EN & 1,000 & 0.485 & 0.483 & 0.002 [$-$0.008, 0.012] & 0.328 \\
\bottomrule
\end{tabularx}
\end{table*}

\subsection{Blood: Selection Strategy Determines Feature Efficiency}

\begin{figure}[!htbp]
\centering
\includegraphics[width=1.0\linewidth]{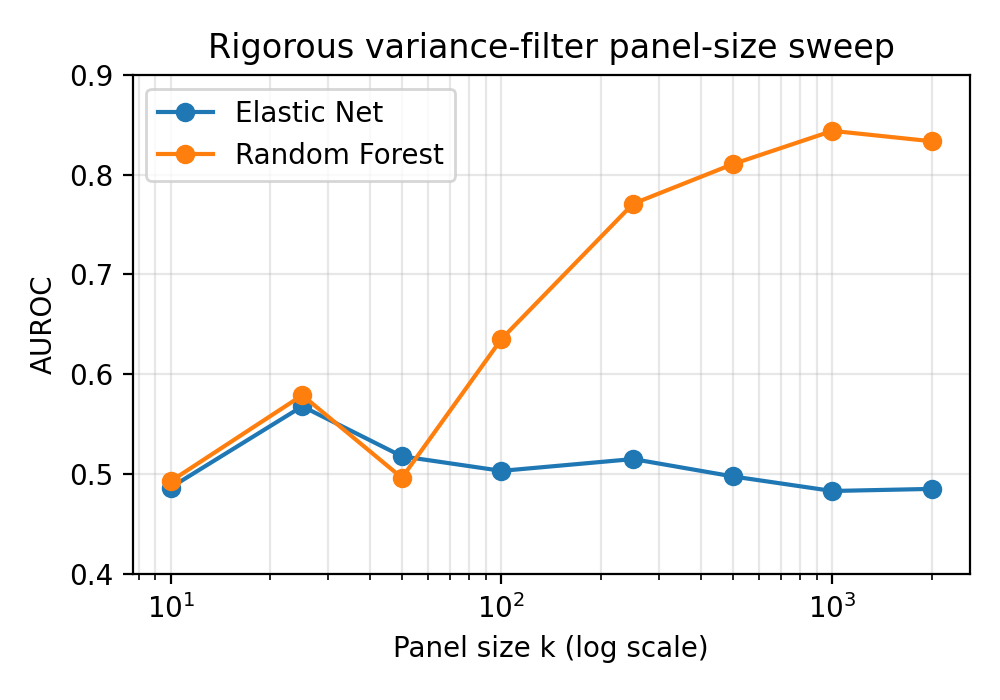}
\caption{Fold-internal variance-filter panel-size sweep: Random Forest improves with panel size while elastic net remains at chance.}
\label{fig:sweep}
\end{figure}

In the fold-internal variance sweep (Table \ref{tab:sweep}, Fig. \ref{fig:sweep}), elastic net remained near chance (0.48--0.57) at every panel size, whereas Random Forest improved steadily with panel size to 0.844 at $k=$ 1,000.

\begin{table}[!htbp]
\centering
\caption{Fold-internal variance-filter panel-size sweep (mean AUROC over 15 folds)}
\label{tab:sweep}
\small
\renewcommand{\arraystretch}{1.15}
\begin{tabular}{ccc}
\toprule
\textbf{Panel size k} & \textbf{Elastic net} & \textbf{Random Forest} \\
\midrule
10 & 0.487 & 0.493 \\
25 & 0.568 & 0.579 \\
50 & 0.518 & 0.497 \\
100 & 0.503 & 0.635 \\
250 & 0.515 & 0.771 \\
500 & 0.498 & 0.810 \\
1,000 & 0.483 & 0.844 \\
2,000 & 0.485 & 0.833 \\
\bottomrule
\end{tabular}
\end{table}

\begin{figure}[!htbp]
\centering
\includegraphics[width=1.0\linewidth]{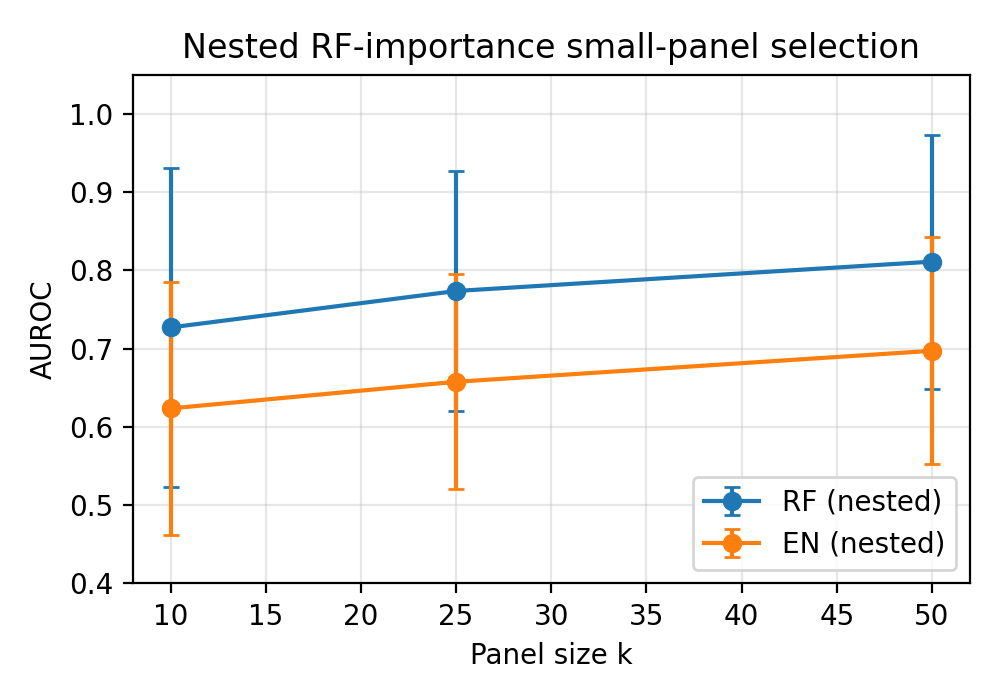}
\caption{Nested importance-based panel selection: a 50-probe panel approaches the 1,000-probe reference, and nested selection lifts elastic net well above its variance-selected counterpart.}
\label{fig:nested}
\end{figure}

Selecting probes by nested Random Forest importance changed the picture for small panels (Table \ref{tab:nested}, Fig. \ref{fig:nested}). With 50 probes, Random Forest reached 0.811 and elastic net 0.697, compared with 0.497 and 0.518 for 50 variance-selected probes. Against the 1,000-probe Random Forest reference on identical folds (0.844 versus 0.811), the difference was 0.033 (95\% CI [$-$0.065, 0.130], corrected $p$ = 0.486): a 20-fold reduction in panel size with no detectable loss of discrimination. The one-sided upper bound on any loss was 0.113, so a formal non-inferiority claim at a 0.05 margin is a natural target for a larger cohort.

\begin{table}[!htbp]
\centering
\caption{Nested Random-Forest-importance panel selection (mean $\pm$ SD AUROC over 15 folds)}
\label{tab:nested}
\small
\renewcommand{\arraystretch}{1.15}
\begin{tabular}{lcc}
\toprule
\textbf{Panel size k} & \textbf{Random Forest} & \textbf{Elastic net} \\
\midrule
10 & 0.727 $\pm$ 0.204 & 0.624 $\pm$ 0.162 \\
25 & 0.774 $\pm$ 0.154 & 0.658 $\pm$ 0.137 \\
50 & 0.811 $\pm$ 0.162 & 0.697 $\pm$ 0.145 \\
1,000 (variance reference) & 0.844 $\pm$ 0.152 & 0.483 $\pm$ 0.062 \\
\bottomrule
\end{tabular}
\end{table}

Across the 15 folds, 19 probes were selected in at least half of the folds and 5 in at least 80\% (Appendix \ref{app:panel}). The most stable probes include \emph{SATB2} and \emph{KAT6A}, each selected in 13 of the 15 folds; both are chromatin regulators, consistent with the chromatin-remodelling theme reported for this cohort by Calligaris et al. \cite{calligaris2015blood}.

\subsection{Blood: Recruitment-Wave Structure in GSE72267}

In the full cohort, expression predicted recruitment wave with AUROC 0.855, and 15 of the 50 most diagnosis-predictive probes were also among the 50 most wave-predictive probes. Because the unpaired wave is 22 of 23 PD, the decisive test is the within-patient analysis. Among PD patients alone, expression predicted recruitment wave with AUROC 0.906 (95\% CI [0.759, 1.000]); the permutation statistic was 0.882 against a null mean of 0.515 (95th percentile 0.703), $p$ = 0.010. Expression therefore separates the two recruitment groups even among patients with the same diagnosis, revealing technical or recruitment-related structure in GSE72267 that, to our knowledge, has not been reported. Of the 19 stable panel probes, 3 also rank among the 50 most wave-predictive probes within PD patients, \emph{KAT6A} among them, which identifies the probes whose biological interpretation requires particular care.

\begin{figure}[!htbp]
\centering
\includegraphics[width=1.0\linewidth]{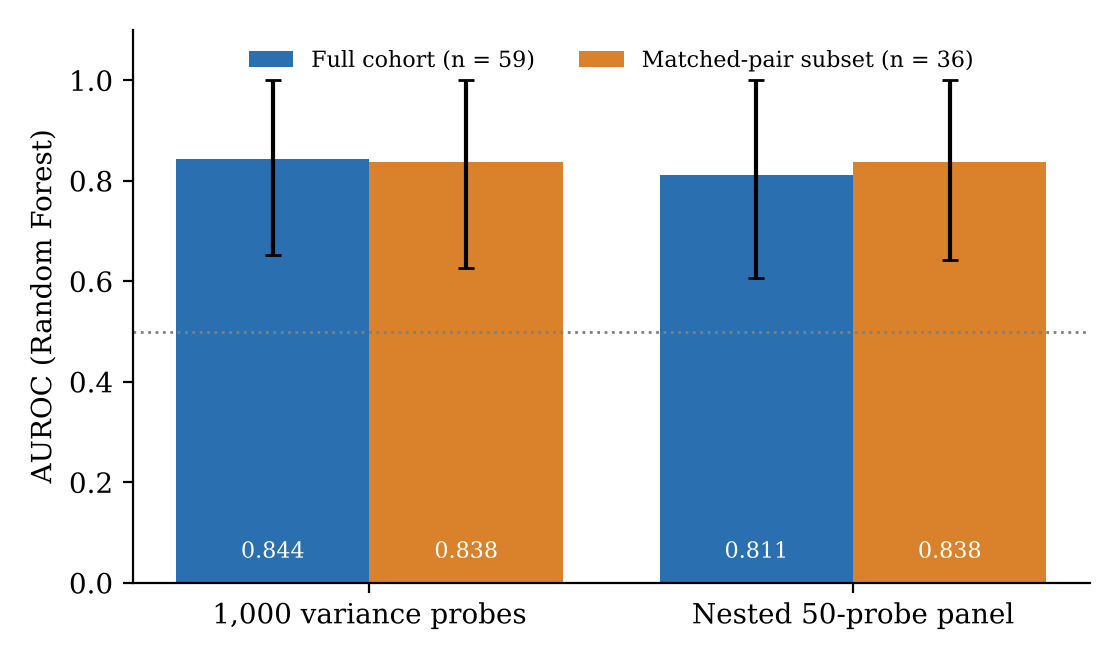}
\caption{Diagnostic AUROC in the full cohort and in the wave-balanced matched-pair subset for the 1,000-probe and nested 50-probe Random Forest. Error bars: corrected 95\% confidence intervals.}
\label{fig:confound}
\end{figure}

Diagnostic signal nevertheless persisted where wave cannot separate the classes. In the wave-balanced matched-pair subset (36 samples, 18 PD and 18 controls), diagnostic AUROC was 0.838 (95\% CI [0.626, 1.000]) with 1,000 probes and 0.838 (95\% CI [0.642, 1.000]) with the nested 50-probe panel (Fig. \ref{fig:confound}), closely matching the full-cohort estimates.

\subsection{Summary of Formal Inference}

\begin{figure}[!htbp]
\centering
\includegraphics[width=1.0\linewidth]{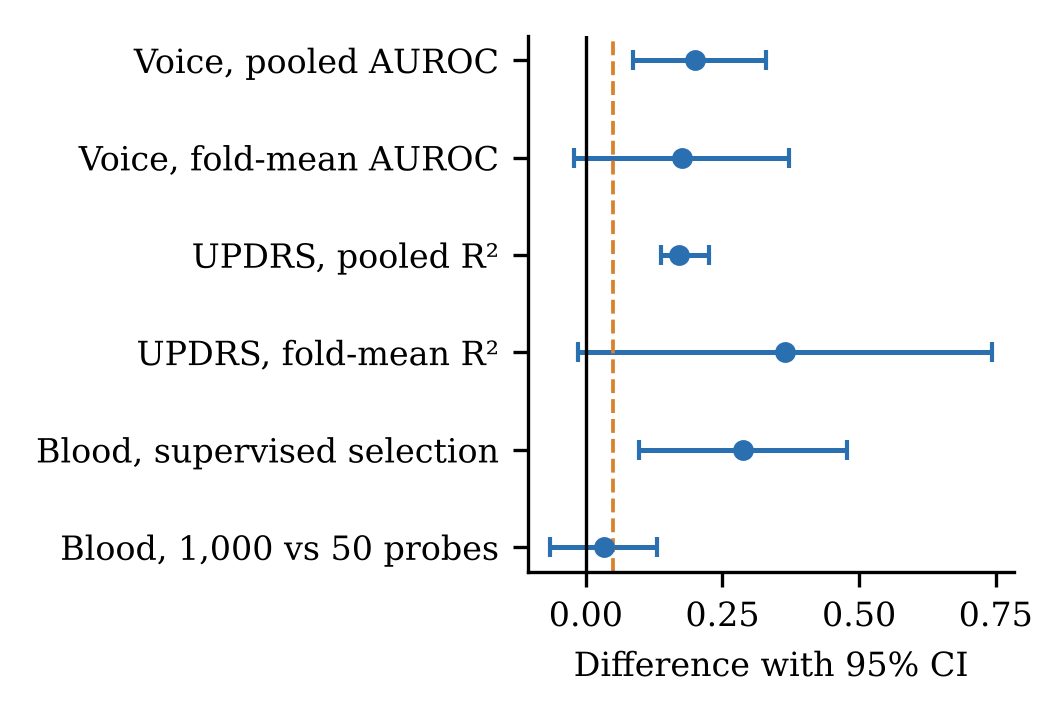}
\caption{Effect sizes (naive minus rigorous, or 1,000-probe minus 50-probe) with 95\% confidence intervals; the dashed line marks a 0.05 AUROC margin.}
\label{fig:forest}
\end{figure}

Table \ref{tab:sig} and Fig. \ref{fig:forest} summarise the headline comparisons under the dependence-aware procedures of Section \ref{sec:stats}.

\begin{table*}[!tb]
\centering
\caption{Headline comparisons with dependence-aware inference. A and B are the naive (or reference) and rigorous (or compact-panel) estimates; p-values for naive-versus-rigorous comparisons are one-sided and confidence intervals two-sided.}
\label{tab:sig}
\footnotesize
\renewcommand{\arraystretch}{1.15}
\begin{tabularx}{\linewidth}{Xccp{3.3cm}cp{1.9cm}}
\toprule
\textbf{Comparison} & \textbf{A} & \textbf{B} & \textbf{Difference [95\% CI]} & \textbf{p} & \textbf{Test} \\
\midrule
Voice AUROC, pooled out-of-fold: record- vs subject-level & 0.892 & 0.693 & 0.200 [0.087, 0.330] & $<$0.001 & subject bootstrap \\
Voice AUROC, fold mean: record- vs subject-level & 0.899 & 0.723 & 0.175 [$-$0.021, 0.371] & 0.039 & corrected t \\
UPDRS $R^2$, pooled out-of-fold: record- vs subject-level & 0.106 & $-$0.065 & 0.171 [0.138, 0.225] & $<$0.001 & subject bootstrap \\
UPDRS $R^2$, fold mean: record- vs subject-level & 0.105 & $-$0.259 & 0.365 [$-$0.014, 0.743] & 0.029 & corrected t \\
Blood AUROC, supervised selection: full cohort vs in fold ($k=$ 50, EN) & 0.998 & 0.711 & 0.287 [0.096, 0.477] & 0.003 & corrected paired t \\
Blood AUROC, 1,000-probe vs nested 50-probe RF & 0.844 & 0.811 & 0.033 [$-$0.065, 0.130] & 0.486 & corrected paired t \\
\bottomrule
\end{tabularx}
\end{table*}

\section{Discussion}\label{sec:discussion}

Subject-level leakage generalises across tasks; therefore, the first research question is answered affirmatively. Both repeated-measures datasets show the predicted pattern despite differing in task, metric and repeated-measures intensity. In the voice data, record-level evaluation overstated AUROC by 0.11 to 0.20 depending on the protocol pair, and the paired subject-level bootstrap places the difference well above zero. The voice signal does not disappear under subject-aware evaluation: an AUROC of about 0.69 to 0.78 remains, so the published voice result is best described as inflated rather than spurious, in agreement with the direction reported by Tougui et al. \cite{tougui2021impact}. In the telemonitoring data the effect is more severe: a modest record-level $R^2$ becomes no better than predicting the mean for a new patient, and the dominance of age indicates that the record-level model largely learned patient identity. Neither Random Forest result rescued performance, so the inflation is a property of the protocol, not of the linear model. This extends prior PD leakage evidence from classification to longitudinal severity regression, where its consequences are larger.

Supervision is the leakage mechanism in transcriptomic analysis, a finding that addresses the second research question and yields the largest effect in the study. Variance filtering on the full cohort does not use labels and produced no inflation, whereas ranking probes by their association with diagnosis on the full cohort drove a model that is otherwise informative at 0.71 to near-perfect apparent accuracy. This reproduces the mechanism described by Ambroise and McLachlan \cite{ambroise2002selection} in a PD cohort and shows that the safeguard that matters is not 'avoid pre-filtering' but 'never let labels reach a selection step outside the training fold'. It is directly relevant to blood-signature studies that select differentially expressed genes before cross-validation \cite{scherzer2007molecular,han2025blood}.

Feature efficiency depended on selection strategy rather than panel size. Variance-selected panels left elastic net at chance at every size, whereas nested importance-based selection raised both models substantially and allowed 50 probes to match the 1,000-probe reference within sampling error. The result points to a mechanism, predictive importance computed inside the training fold, rather than to an unexplained correlation with feature count, which makes it more actionable for the compact blood panels now being pursued \cite{han2025blood,gao2025multiplex}.

Recruitment structure is a hidden confounder in a public cohort; therefore, the third research question is answered by a design in which batch cannot masquerade as diagnosis. Expression predicted recruitment wave among patients alone well above the permutation null, so GSE72267 carries non-biological structure aligned with recruitment. Because the unpaired wave is almost entirely PD, part of the full-cohort diagnostic signal may reflect this structure; the matched-pair analysis shows that diagnostic signal of similar magnitude remains when the classes are balanced across waves. This finding has practical value for every study that reuses GSE72267, and the within-outcome-class test is a general template for auditing batch structure when batch labels are not recorded.

A key implication for PD machine learning is that, for repeated-measures data, subject-aware splitting should be a minimum standard, and accuracies from record-level splits should be read as upper bounds. For transcriptomic PD data, every supervised step must sit inside the training fold, and cohort structure should be audited with tests that are independent of the outcome. For all PD-ML work, protocol comparisons should use inference that respects fold dependence, and repeated grouped cross-validation should be verified to produce distinct partitions. Interpretability outputs such as SHAP inherit the properties of the evaluation design, so they are most informative when computed on leakage-free models \cite{dentamaro2024ppmi}.

\section{Evaluation of the Research Method}\label{sec:evaluation}

The within-dataset controlled comparison proved an effective way to answer the research questions. Holding the model family and tuning fixed produced interpretable inflation deltas, and fitting Random Forest on identical partitions separated protocol from capacity directly. The leakage-only ablation confirmed that grouping alone accounts for the voice effect, and the two-arm design for transcriptomics isolated supervision as the operative mechanism. The paired subject-level bootstrap of pooled out-of-fold predictions was particularly effective: it allowed record-level and subject-level protocols, which never share test folds, to be compared on the same records while respecting within-subject dependence.

Framing the problem around data structure rather than modality made the three datasets complementary: two repeated-measures datasets test subject-level leakage under different tasks, and a $p \gg n$ dataset tests selection leakage and cohort structure. The literature on selection bias and fold dependence \cite{ambroise2002selection,varma2006bias,cawley2010overfitting,nadeau2003inference} shaped the analytical choices, in particular the separation of supervised from unsupervised selection and the use of corrected inference. Generating every reported number programmatically from saved outputs ensured that text, tables and figures remain consistent with the analysis. Table \ref{tab:aims} maps each research objective to the evidence obtained; every objective was addressed by a dedicated analysis.

\begin{table}[!htbp]
\centering
\caption{Achievement of the research objectives}
\label{tab:aims}
\small
\renewcommand{\arraystretch}{1.15}
\begin{tabularx}{\linewidth}{p{2.7cm}X}
\toprule
\textbf{Objective} & \textbf{Evidence} \\
\midrule
Leakage in repeated-measures data & Voice and UPDRS inflation deltas; leakage-only ablation; paired subject-level bootstrap. \\
Role of model capacity & Random Forest on identical partitions does not close either gap. \\
Pre-split selection & Supervised arm inflates AUROC by about 0.2--0.3; unsupervised arm does not. \\
Compact panel & 50 probes match the 1,000-probe reference within sampling error; a stable core of 19 probes identified. \\
Recruitment structure & Within-PD wave AUROC well above permutation null; diagnostic signal retained in matched pairs. \\
Dependence-aware inference & Corrected $t$-tests, distinct subject partitions and subject-level bootstrap used for all headline comparisons. \\
\bottomrule
\end{tabularx}
\end{table}

The limitations of the study, and the recommendations that follow from its results, are as follows. 

Each conclusion rests on a single public cohort per modality, the voice and blood control groups are small (8 and 19), and recruitment waves are inferred from file names rather than recorded metadata. These boundaries define the natural next steps rather than qualifying the main findings, which are consistent across protocols, models and inference procedures. We recommend that future PD-ML studies (i) use group-aware splitters with explicit shuffling and verify that repeats yield distinct partitions; (ii) report pooled out-of-fold estimates with subject-level resampling; (iii) include a supervised-selection arm whenever pre-split selection is examined; (iv) pre-specify equivalence margins for panel-reduction claims; and (v) audit batch structure within outcome classes.

\section{Conclusion}\label{sec:conclusion}

This study asked how much of the performance reported for Parkinson's disease prediction models reflects the evaluation protocol, and whether the answer depends on data structure. The answer is that protocol effects are large, that they take different forms in different data structures, and that each form has a specific safeguard.

\begin{itemize}
\item \textbf{Subject-level leakage materially inflated performance in both repeated-measures datasets.} Voice AUROC fell from 0.892 to 0.780 under subject-aware evaluation, and the leakage-only ablation gave a pooled paired difference of 0.200 (95\% CI [0.087, 0.330]). Telemonitoring $R^2$ fell from 0.105 to $-$0.037, no better than predicting the mean for an unseen patient.
\item \textbf{Model capacity did not recover the lost performance.} On identical subject-aware partitions, Random Forest reached AUROC 0.829 for voice, still below the naive estimate, and $R^2$ of $-$0.483 for UPDRS, lower than elastic net. The inflation is a property of the protocol, not of the linear model.
\item \textbf{In transcriptomics, supervised selection, not pre-filtering, is the leakage mechanism.} Variance filtering on the full cohort produced no inflation, whereas ranking probes by their association with diagnosis on the full cohort raised elastic-net AUROC from 0.711 to 0.998 ($p$ = 0.003).
\item \textbf{Selection strategy, not panel size, determined feature efficiency.} Nested importance-based selection reached AUROC 0.811 with 50 probes against 0.844 with 1,000, a 20-fold reduction with no detectable loss, whereas 50 variance-selected probes performed at chance.
\item \textbf{GSE72267 contains recruitment-wave structure not reported in the original study.} Expression predicted recruitment wave among PD patients alone with AUROC 0.906 (permutation $p$ = 0.010), yet diagnostic signal persisted in the wave-balanced matched-pair subset (AUROC 0.838 with 1,000 probes and 0.838 with 50).
\item \textbf{Valid protocol comparison requires dependence-aware inference.} Row shuffling does not re-randomise GroupKFold, and fold scores from repeated cross-validation are dependent; distinct subject partitions, corrected $t$-tests and a paired subject-level bootstrap provide reliable comparisons.
\end{itemize}

The contribution of this work is methodological: it shows directly how evaluation choices change conclusions about PD prediction, that different data structures fail in different ways, and which safeguard each requires. Future work will validate a stability-selected probe panel on an independent cohort with recorded batch information and nested batch correction, perform pathway enrichment on the stable probes, examine the contribution of age in the telemonitoring task, and extend the audit framework to multimodal PPMI data \cite{dentamaro2024ppmi,makarious2022multimodality}.

\appendices

\section{Most Frequently Selected Probes in the Nested 50-Probe Panel}\label{app:panel}

Probes selected in at least half of the 15 folds, with GPL571 gene symbols. Freq. is the selection frequency across folds; Wave (PD) marks membership of the 50 most wave-predictive probes among PD patients.

\begin{table}[!htbp]
\centering
\caption{Stable probes of the nested 50-probe panel}
\label{tab:panel}
\footnotesize
\renewcommand{\arraystretch}{1.15}
\begin{tabular}{llcc}
\toprule
\textbf{Probe} & \textbf{Gene} & \textbf{Freq.} & \textbf{Wave (PD)} \\
\midrule
\texttt{219840\_s\_at} & \emph{TCL6} & 0.93 &  \\
\texttt{221753\_at} & \emph{SSH1} & 0.93 & yes \\
\texttt{213435\_at} & \emph{SATB2} & 0.87 &  \\
\texttt{216361\_s\_at} & \emph{KAT6A} & 0.87 & yes \\
\texttt{212113\_at} & \emph{ATXN7L3B} & 0.80 &  \\
\texttt{201340\_s\_at} & \emph{ENC1} & 0.73 &  \\
\texttt{204589\_at} & \emph{NUAK1} & 0.73 &  \\
\texttt{217207\_s\_at} & \emph{BTNL3} & 0.73 &  \\
\texttt{209789\_at} & \emph{CORO2B} & 0.73 &  \\
\texttt{212743\_at} & \emph{RCHY1} & 0.73 &  \\
\texttt{206108\_s\_at} & \emph{SRSF6} & 0.67 &  \\
\texttt{211663\_x\_at} & \emph{PTGDS} & 0.67 &  \\
\texttt{217671\_at} & \emph{DSERG1} & 0.67 &  \\
\texttt{222285\_at} & \emph{IGHD} & 0.60 &  \\
\texttt{215390\_at} & -- & 0.60 &  \\
\texttt{214464\_at} & \emph{CDC42BPA} & 0.60 & yes \\
\texttt{209841\_s\_at} & \emph{LRRN3} & 0.60 &  \\
\texttt{212187\_x\_at} & \emph{PTGDS} & 0.53 &  \\
\texttt{203470\_s\_at} & \emph{PLEK} & 0.53 &  \\
\bottomrule
\end{tabular}
\end{table}

\end{document}